\documentclass[twocolumn,aps,prl,showpacs,amsmath,amssymb,nobalancelastpage]{revtex4}
\usepackage[dvips]{graphicx}
\usepackage{bm,eucal}
\begin{document}
%
%
\newcommand{\nn}{\nonumber\\}
\newcommand{\rbr}[1]{\left(#1\right)}
\newcommand{\sbr}[1]{\left[#1\right]}
\newcommand{\cbr}[1]{\left\{#1\right\}}
\newcommand{\av}[1]{\prec #1 \succ}
\newcommand{\eq}[1]{(\ref{#1})}
\newcommand{\fou}[1]{\check{#1}}                           
\newcommand{\fvol}[2]{\frac{d^{#2} #1}{(2\,\pi)^{#2}}}    
\newcommand{\de}{\partial}                                         
\newcommand{\der}[2]{\frac{d #1}{d #2}}               
\newcommand{\pder}[2]{\frac{\partial #1}{\partial #2}} 
\newcommand{\fde}[2]{\frac{\delta #1}{\delta #2}}      
\newcommand{\fset}[2]{\chi_{#1}\left(#2\right)}   
\newcommand{\dirac}[2]{\delta^{(#1)}(#2)}         
\newcommand{\flu}[1]{\delta #1}                   
\newcommand{\vv}[1]{\mathbf{#1}}                  
\newcommand{\uv}[1]{\hat{\mathbf{#1}}}                  
\newcommand{\fphi}{\check{\phi}}
\newcommand{\bj}{\bar{\jmath}}
\newcommand{\bv}{\bar{v}}
\newcommand{\fv}{\fou{v}}
\newcommand{\fbv}{\fou{\bar{v}}}
\newcommand{\eps}{\varepsilon}
\newcommand{\mf}{m_{F}}
\newcommand{\dd}[1]{\eta_{#1}}
\newcommand{\ac}{\mathcal{A}}
\newcommand{\partfun}{\mathcal{Z}}
\newcommand{\mes}{\mathcal{D}}
\newcommand{\stf}[1]{\mathcal{S}_{#1}}
\newcommand{\cf}{\mathcal{C}}
\title{On the scaling properties of $2d$ randomly stirred Navier--Stokes equation}
\author{Andrea Mazzino}
\email{mazzino@fisica.unige.it}
\affiliation{Department of Physics, University of Genova, and  INFN, CNISM, Genova Section, 
Via Dodecaneso 33, I--16146 Genova, Italy,}
\author{Paolo Muratore-Ginanneschi}
\email{paolo.muratore-ginanneschi@helsinki.fi}
\affiliation{Department of Mathematics and Statistics, University of Helsinki, 
P.O. Box 4, 00014 Helsinki, Finland,}
\author{Stefano Musacchio}
\email{stefano.musacchio@gmail.com}
\affiliation{Department of Physics of Complex Systems
Weizmann Institute of Science
Rehovot 76100, Israel
} 
\date{\today}
\begin{abstract}
We inquire the scaling properties of the $2d$ Navier-Stokes equation
sustained by a forcing field with Gaussian statistics, white-noise in time and 
with power-law correlation in momentum space of degree $2-2\,\eps$. This is at variance 
with the setting usually assumed to derive Kraichan's classical theory. We contrast accurate 
numerical experiments with the different predictions provided  for the small $\eps$ regime 
by Kraichnan's double cascade theory and by renormalization group (RG) analysis. 
We give clear evidence that for all $\eps$ Kraichnan's theory is consistent with the observed 
phenomenology. Our results call for a revision in the RG analysis of ($2d$) fully 
developed turbulence.
\end{abstract}
\maketitle

In two dimensions, the joint conservation of energy and enstrophy 
has relevant consequences for the Navier--Stokes equation.
In $2d$ it is possible to prove that in the deterministic case the solution of 
the Cauchy problem 
exists and is unique \cite{Lad69} and, very recently, 
that in the stochastic case \cite{KS00, BLK00, EMS01, BLK01, BLK02, Mat02} 
the solution is a Markov process exponentially mixing in time 
and ergodic with a unique invariant (steady state) measure 
even when the forcing acts only on two Fourier modes \cite{HM04}. 
At large Reynolds number, the $2d$ scaling properties also differ from the $3d$-case. 
A long standing hypothesis \cite{Pol92},  
very recently corroborated by numerical experiments \cite{BeBoCeFa06,BeBoCeFa06b}, 
surmises the existence of a conformal invariance in $2d$. 
A phenomenological theory due to Kraichnan \cite{Kr67} and Batchelor \cite{Bat69} 
predicts the presence of a double cascade mechanism governing the transfer 
of energy and enstrophy in the limit of infinite inertial range. 
Accordingly, an \emph{inverse energy cascade} with spectrum 
characterized by a scaling exponent $d_{\mathcal{E}}=-5/3$ appears for values of the 
wave-number $k$ smaller than the typical scale $k_{F}$ of the forcing scale.  
For wave-numbers larger than $k_{F}$ a \emph{direct enstrophy cascade} should occur. 
The corresponding energy spectrum scales as  $d_{\mathcal{E}}=-3+\dots$ 
where the dots here stand for possible logarithmic corrections.
As emphasized in \cite{Be99,Be00} Kraichnan's theory is encoded in three hypothesis, (i) 
velocity correlations are smooth at finite viscosity and exist in the inviscid limit even at 
coinciding points, (ii) Galilean invariant functions and in particular structure
functions reach a steady state and (iii) no dissipative anomalies occur for the energy cascade.
Under these hypotheses, if  the forcing field is homogeneous and isotropic Gaussian 
and time $\delta$-correlated it is possible to derive asymptotic expressions 
of the three point structure functions of the velocity field 
consistent with Kraichnan's predictions \cite{Be99,Be00,Lindborg}.
Very strong laboratory (see \cite{KeGo02,Ta02} and reference therein) and numerical evidences 
(see e.g. \cite{Bo06} and references therein) support Kraichnan' s theory. 
However, a first principle derivation of the statistical properties 
of two dimensional turbulence is still missing.
An attempt in this direction has recently been undertaken \cite{Ho98,Ho02} 
(see also \cite{Olla91} and \cite{AdHoKoVa05}) by applying a renormalization group 
improved perturbation theory (RG) (\cite{Zinn} and \cite{AdAnVa} 
for review of applications to fluid turbulence) to the randomly stirred Navier--Stokes equation 
with power law forcing. However how it will be discussed in details in the sequel, RG analysis
leads to a scenario not a priori consistent with Kraichnan's theory. The goal of the present work
is to shed light on this issue. 

RG starting point is the randomly stirred Navier--Stokes equation
\begin{eqnarray}
&&(\de_{t}+v\cdot\de)v^{\alpha}-\nu_{0}\,\de^2 v^{\alpha}=-\de^{\alpha}P 
+ f^{\alpha}-\xi_{0}\,v^{\alpha}
\label{NS}
\\
&& \de\cdot v=0 \nonumber
\end{eqnarray}
where $\alpha=1,2$,  $P$ is the pressure enforcing incompressibility 
and $\xi_{0}$ is an Eckman type coupling providing for large scale dissipation. 
The forcing $f$ is a Gaussian field with zero average 
and correlation
\begin{eqnarray}
&&\av{f^{\alpha}(\vv{x},t)f^{\beta}(\vv{y},s)}=\delta(t-s)\,F^{\alpha\,\beta}(\vv{x}-\vv{y})
\\
&&F^{\alpha\,\beta}(\vv{x}):= \int\fvol{p}{d}\,e^{\imath \vv{p}\cdot \vv{x}}\,\fou{F}(p)\,
T^{\alpha\,\beta}(\uv{p})
\label{correlation}
\\
&&\fou{F}(p):=\frac{F_{0}\,\chi\,\rbr{\frac{p}{m},\frac{p}{M}}}{p^{d-4+2\,\eps}}
\label{correlation:core}
\end{eqnarray}
$F_{0}$ is a constant specifying the amplitude of forcing fluctuations 
and $T^{\alpha\,\beta}(p)$ is the transversal projector.
The function $\chi$ in \eq{correlation:core}  is slowly varying 
for $m\ll p \ll M$ and set infra-red $m$ and ultra-violet cut-offs $M$ scales for the forcing. 
Its detailed shape does not affect the scaling predictions of the RG.
Irrespectively of the spatial dimension $d$,  
the cumulative spectrum $F^{\alpha}_{\,\,\,\alpha}(0)$ of the forcing 
(Einstein convection for repeated vector indexes) 
diverges  for $0\,\leq\, \eps\, <\, 2$ as the ultra-violet cut-off $M$ tends to infinity 
hence providing for stirring at small spatial scales. 
For $\eps\,>\,2$,  $F^{\alpha}_{\,\,\,\alpha}(0)$ is dominated by small wave-numbers 
and thus describes infra-red stirring.  
At $\eps=0$, the scaling dimensions of the material derivative, 
dissipation and forcing in \eq{NS} coincide 
for a scaling dimension of the velocity field $d_{v}=1$ in momentum units. 
Thus $\eps=0$ provides a marginal limit around which scaling dimensions 
can be perturbatively determined with the help of ultra-violet RG. 
The main result \cite{Ho98,Ho02,AdHoKoVa05} is the existence of a 
non-Gaussian infra-red stable fixed point of the RG flow yielding 
for the energy spectrum the prediction
\begin{eqnarray}
\mathcal{E}(k)= \eps^{1/3} \,F_{0}^{2/3}\,k^{1-\frac{4\,\eps}{3}}\, R\rbr{\eps, 
\frac{m}{k},\frac{k_{b}}{k}}
\label{RGspectrum}
\end{eqnarray}
The adimensional function $R$ depends upon infra-red scales $m$ 
and $k_{b}=(\eps\, \xi_{0}^{3}/ F_{0})^{1/(6-2\eps)}$ 
and admits a regular expansion in powers of $\eps$ at the RG fixed point \cite{Ho98}. 
The resummation leading to $k^{1-\frac{4\,\eps}{3}}$ in \eq{RGspectrum} is derived from
the solution of the Callan--Symanzik equation \cite{Zinn} which in the present case requires 
scaling at finite $\eps$ to stem from the balance of the two terms in the material derivative 
with the forcing i.e. from the requirement of Galilean invariance alone.   
According to \cite{AdAnVa,Ho98,Ho02} composite operators at small $\eps$ 
does not induce any self-similarity breaking by the infra-red scales $m$ and $k_{b}$.  
The conclusion is that the spectrum should scale for small $\eps$ 
with exponent $d_{\mathcal{E}}=1-4\eps/3$ as in the $3\,d$-case \cite{AdAnVa}.
Such conclusion seems at variance with Kraichnan's theory which instead 
suggests the occurrence of an inverse cascade at small $\eps$. 
Namely  under the assumptions (i),(ii),(iii) \cite{Be99,Be99b} the energy balance equation
\begin{eqnarray}
\lefteqn{\rbr{\de_{t}+\xi_{0}-\nu_{0}\de^{2}}\av{v^{\alpha}(\vv{x},t)v_{\alpha}(0,t)}}
\nonumber\\
&&-\frac{1}{2}\de_{\mu}\av{||\flu{v}||^{2}(\vv{x},t)\,\flu{v}^{\mu}(\vv{x},t)}=F(\vv{x})
\label{enbal}
\end{eqnarray}
with $\flu{\vv{v}}^{\alpha}(\vv{x},t)=v^{\alpha}(\vv{x},t)-v^{\alpha}(0,t)$  
yields for $\xi_{0}=0$, $\eps<2$ and $r^2:=||x||^2$ 
\begin{eqnarray}
S_{3}(r)\simeq c_{1}F_{0}M^{4-2\eps}r+\frac{c_{2}\,F_{0}}{r^{3-2\eps}}\,,\quad  
m^{-1}\gg\,r\,\gg M^{-1} 
\label{s3}
\end{eqnarray}
$S_3$ denotes the three point velocity longitudinal structure function and $c_{i}$ $i=1,2$ two 
adimensional coefficients irrelevant for the present argument. 
Power counting based on \eq{s3} then predicts a $d_{\mathcal{E}}=-5/3$ inverse cascade spectrum 
at small $\eps$ with \emph{at most sub-leading} corrections consistent with the RG  prediction. 

\begin{figure} [h!]
\centerline{
\includegraphics[scale=0.6,draft=false]{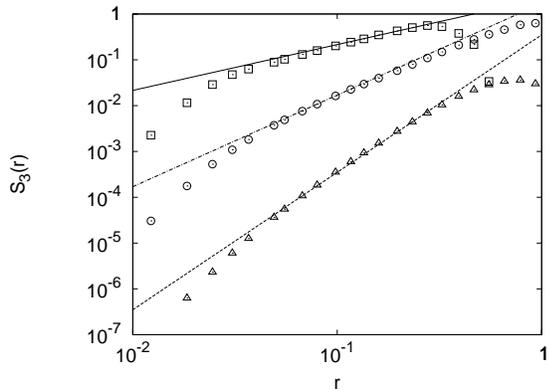}}
\caption{Third order structure function of 
longitudinal velocity increments $S_3(r)$ for 
$\epsilon=1$ (squares), 
$\epsilon=2.5$ (circles), 
$\epsilon=4$ (triangles). 
When $\epsilon<2$ the presence of an inverse energy cascade 
with constant flux results in the scaling $S_3(r)\sim r$ (solid line).
When $\epsilon>3$ the presence of a direct enstrophy cascade 
with constant flux results in the scaling $S_3(r)\sim r^3$ (dashed line).
For $2<\epsilon<3$ the scaling is compatible with the RG argument (dash-dotted line)
}
\label{fig:2}
\end{figure}

\begin{figure} [h!]
\includegraphics[scale=0.5,draft=false]{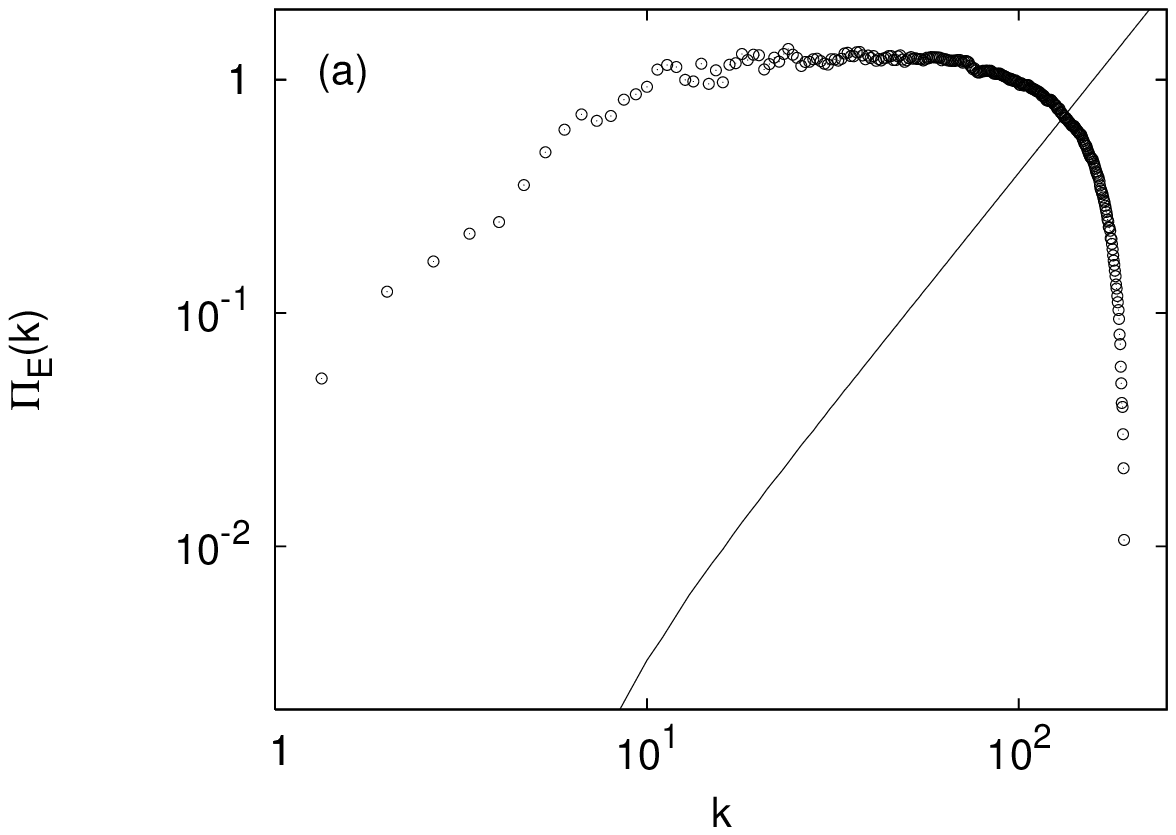}
\includegraphics[scale=0.5,draft=false]{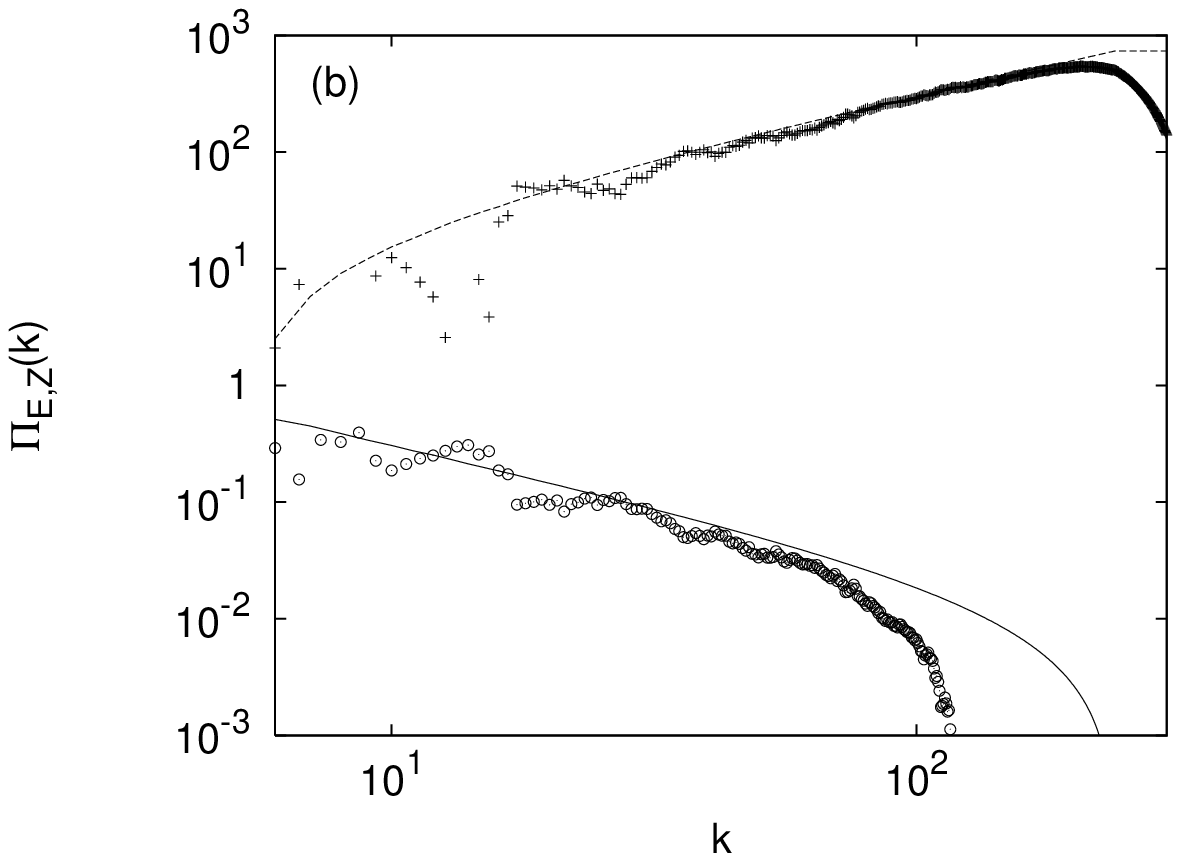}
\includegraphics[scale=0.5,draft=false]{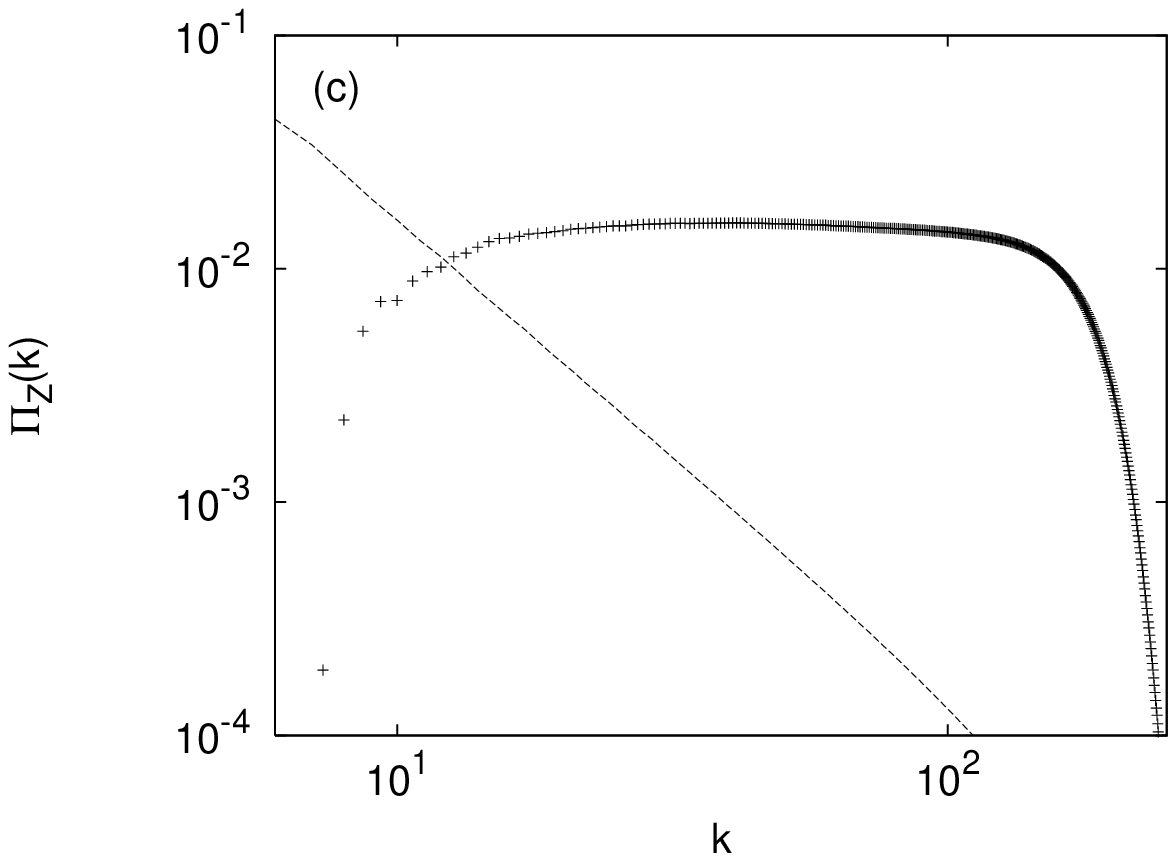}
\caption{
The energy flux $\Pi_{E}$ (symbols) and the energy input $I_{E}(k)$ 
(thin line) for different values of $\epsilon$.
(a) $\epsilon=1$; (b) $\epsilon=2.5$; (c) $\epsilon=4$.
The energy input is defined as
$
I_{E}(k)\propto\int_{0}^{k}dq\,q\, \fou{F}_{v}(q)
$  for $\epsilon< 2$ and as
$
I_{E}(k)\propto\int_{k}^{\infty}dq\,q\,\fou{F}_{v}(q)
$  for $\epsilon > 2$.
In (b) the enstrophy input, 
$I_{Z}(k)\propto \int_{0}^{k}dq\,q^3\,\fou{F}(q)$ 
with and the enstrophy flux $\Pi_{Z}(k)$ are also plotted 
in order to show their balance at each inertial range scale.}
\label{fig:3}
\end{figure}

The currently available numerical resources permit to contrast the two apparently 
discordant predictions \eq{RGspectrum} and \eq{s3} with the actual 
Navier--Stokes phenomenology.  
To attain this goal we integrated the Navier--Stokes equation (\ref{NS})
for the vorticity field ($\omega =\epsilon_{\alpha\,\beta}\de^{\beta} v^{\alpha} $) 
with a standard, fully-dealiased pseudospectral method in a doubly periodic square 
domain of resolution $1024^2$.  
\begin{figure} [h!]
\centerline{
\includegraphics[scale=0.6,draft=false]{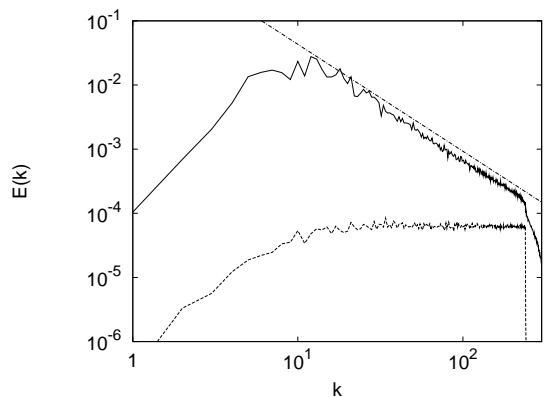}}
\caption{Energy spectrum for $\eps=0.5$ and standard molecular dissipation. 
An inertial range with inverse energy cascade sets in for sufficiently 
small viscosity $\nu=O(10^{-4})$ (solid line). At larger viscosity the inertial range 
is suppressed and dissipative spectrum $\mathcal{E}\sim k^{1-2\,\eps}$ 
is observed (dashed line).}
\label{fig:1}
\end{figure}
Time evolution was computed by means of a standard second-order
Runge--Kutta scheme. 
We repeated our numerical experiments for different values 
of the  hyperviscosity $(-1)^{p+1}\nu_{0}\de^{2\,p}v$ including  $p=1$, 
standard viscosity as in \eq{NS}, and $p=4,6$ obtaining qualitatively identical results.
The outcomes evince that for \emph{all} $0\leq \eps \leq 2$
and sufficiently small viscosity, 
an inverse energy cascade with exponent consistent with $d_{\mathcal{E}}=-5/3$  
is observed (see Fig.~\ref{fig:1}).  
If the viscosity increases,  finite resolution effects prevent 
the observation  of an inertial range as the Kolmogorov scale becomes 
of the order of the infra-red dissipation scale.  
In such a case a dissipative spectrum appears with scaling exponent 
consistent with the prediction $d_{\mathcal{E}}=1-2\,\eps$ dictated 
by the balance between forcing and dissipation (see again Fig.~\ref{fig:1}). 
In agreement with \eq{s3}, local balance scaling 
becomes dominant in the range $2<\eps<\,3$ (see Fig. \ref{fig:2}). 
There, energy spectra scale with an exponent compatible with $d_{\mathcal{E}}=1-4\eps/3$. 
Finally, for $\eps> 3$ a direct enstrophy cascade invades the whole inertial range. 
The phenomenology just described  is confirmed by the inspection 
of the energy and enstrophy fluxes in the inertial range (see Fig.~\ref{fig:3}). 
They are respectively defined by
$
\Pi_{E}(k)\propto \int_{k_{o}}^{k}\fvol{q}{2} 
\Re\av{\mathcal{F}\cbr{v}^{\alpha}(-q)\mathcal{F}\cbr{(v\cdot\de v_{\alpha})}(q)}
$
and
$
\Pi_{Z}(k)\propto \int_{k_{o}}^{k} \fvol{q}{2} 
\Re\av{\mathcal{F}\cbr{\omega}(-q)\mathcal{F}\cbr{(v\cdot\de \omega)}(q)}
\label{essp}
$ where $\mathcal{F}$ denotes the Fourier transform and $k_{o}$ 
is chosen such that $\Pi_{E}(k)$ and $\Pi_{Z}(k)$ are positive quantities in the inertial range.
As spectra and structure functions, fluxes highlight the existence 
versus $\eps$ of three distinct scaling regimes the origin 
of which can be understood by contrasting \eq{enbal} with the 
energy and enstrophy inputs 
$I_{E}(k)\propto\int_{k_{o}}^{k}dq\,q\,\fou{F}_{v}(q)$ and 
$I_{Z}(k)\propto\int_{k_{o}}^{k}dq\,q\,\fou{F}_{v}(q)$.
For $0\leq \eps <\,2$ both $I_{E}$ and  $I_{Z}$ are
peaked in the ultra-violet. Since stirring occurs mainly at small spatial scales  
an inverse energy cascade sets in the whole inertial range.  
In the range $2\,<\,\eps\,<\,3$ the energy input becomes infra-red divergent 
whilst $I_{Z}$ is still ultra-violet divergent in the absence of cut-offs.  
In this situation, the steady state is attained when the energy and enstrophy fluxes 
balance scale by scale the corresponding inputs  (see Fig.~\ref{fig:3}). 
In this $\eps$-range the energy spectrum scales in agreement with the exponent which 
can be extrapolated but no longer fully justified using the RG.  
Finally, for $\eps>3$ both the energy and enstrophy inputs become peaked in the infra-red. 
In such a case \cite{Be99,Be99b} the right hand sides of Eq. \eq{enbal} 
and of the analogous equation for the vorticity correlation 
admit a regular Taylor expansion for $m x \ll 1$. 
The hypotheses (i), (ii), (iii) thus recover  $S_{3}(x)\propto r^3$ 
i.e. Kraichnan's scaling for the \emph{direct enstrophy cascade} 
in agreement with our numerical observations. 

\begin{figure} [h!]
\includegraphics[scale=0.6,draft=false]{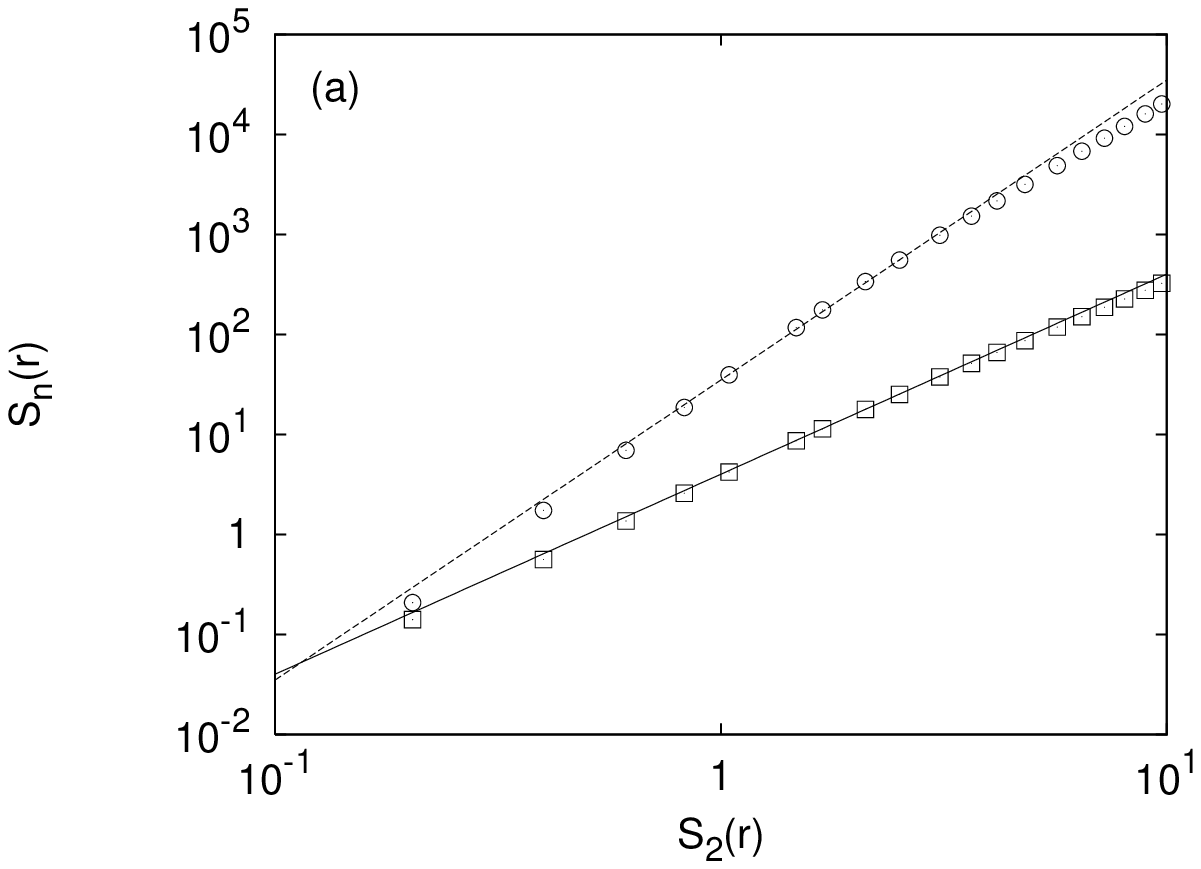}
\includegraphics[scale=0.6,draft=false]{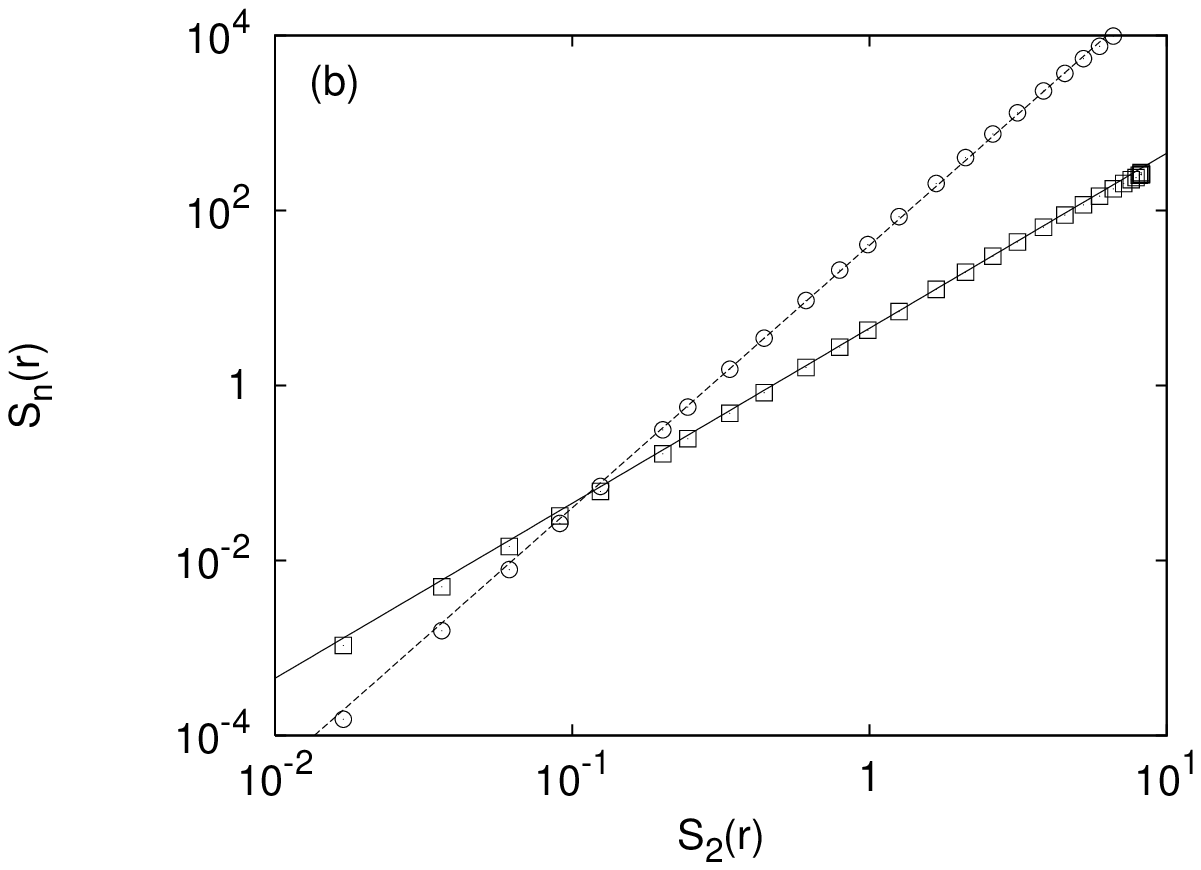}
\includegraphics[scale=0.6,draft=false]{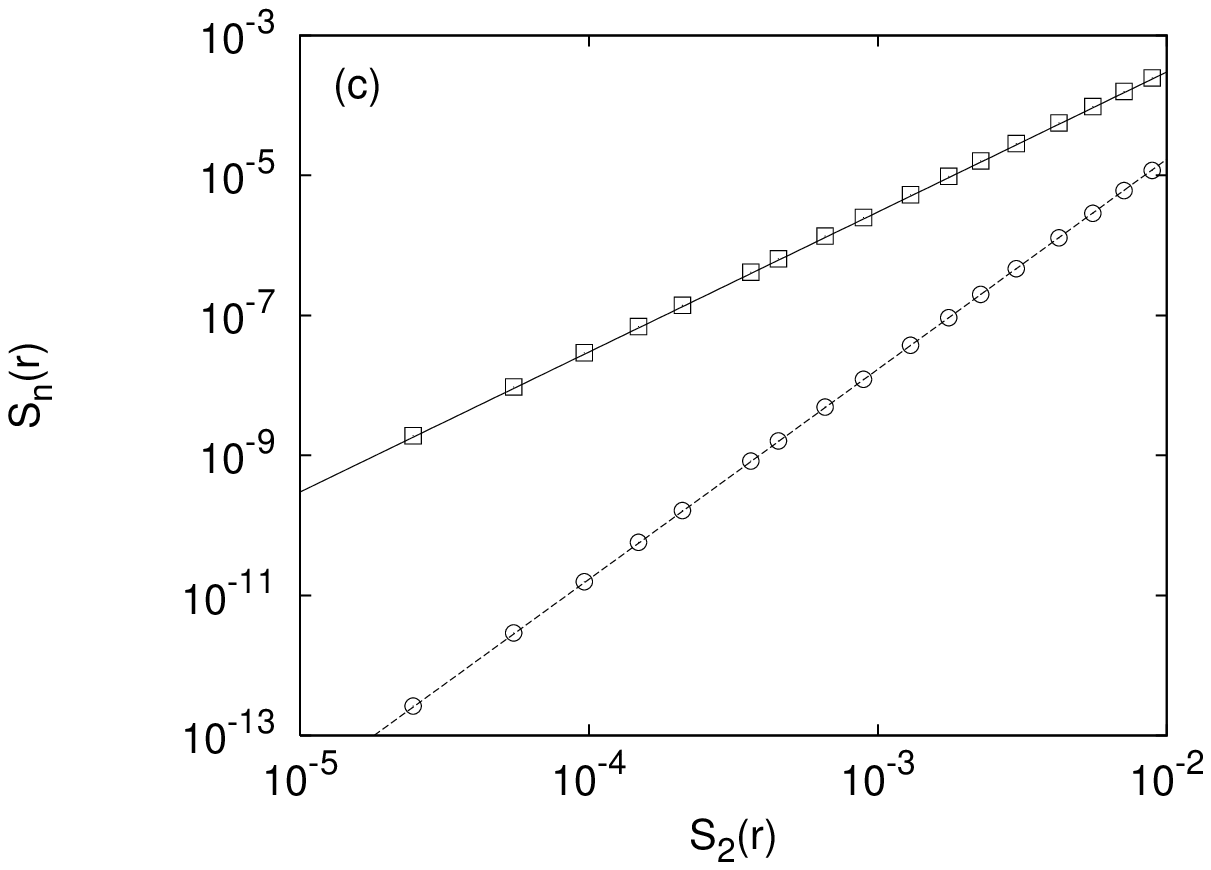}
\caption{
Structure function of longitudinal velocity increments 
$S_4(r)$ (squares) and 
$S_6(r)$ (circles) vs. $S_2(r)$. 
The lines represent the non-intermittent scalings
$S_4(r) \sim S_2(r)^2$ (solid line) and $S_6(r) \sim S_2(r)^3$ (dashed line).
(a) $\epsilon=1$; (b) $\epsilon=2.5$; (c) $\epsilon=4$}
\label{fig:4}
\end{figure}

\begin{figure} [h!]
\centerline{
\includegraphics[scale=0.6,draft=false]{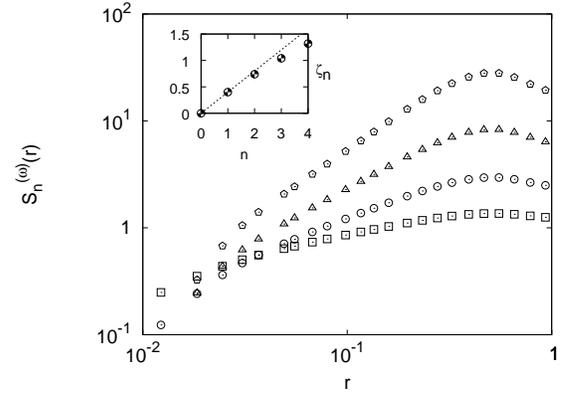}}
\caption{
Structure function of vorticity increments 
$S_n^{(\omega)}(r)$ for $n=1,2,3,4$ (square, circles, triangles, poligons). 
In the inset we show the scaling exponents $\zeta_n$. Here $\epsilon =4$.
}
\label{fig:5}
\end{figure}

As far as intermittency is concerned, our numerics support 
normal scaling of velocity structure functions (Fig.~\ref{fig:4}). 
Higher order vorticity structure function are compatible 
with a weakly anomalous scaling (Fig.~\ref{fig:5}). 
These results provide an indirect positive test for  
the occurrence of a vorticity dissipative anomaly \cite{Be99b}. 

In conclusion, our numerics fully support the validity of Kraichnan's theory 
for all values of the H\"older exponent of power law forcing in $2d$. 
RG scaling prediction is not observed in the range where it was supposed to appear. 
It seems to us that this in not trivially a consequence 
of low Reynolds number ($O(\eps^{1/2})$ for $F_{0}=O(\eps)$ 
see discussion in section 9.6.4 of \cite{Frisch}) entailed by perturbative expansion.
Instead scaling seems to be related to the convolution of the response field
with the forcing kernel in the Schwinger-Dyson equation \cite{AdAnVa,Zinn} giving rise 
to a non-local scaling field of dimension $d_{O}=-1/3$ independently of $\eps$.  
This point however deserves more theoretical inquiry.

We are pleased to thank J.~Honkonen, A.~Kupiainen and P.~Olla 
for numerous friendly and enlightening discussions. 
Numerical simulations have been performed on the ``Turbofarm'' 
cluster at the INFN computing center in Torino.  A.M. was supported by
COFIN 2006 N.2005027808 and by CINFAI consortium, P.M.-G. by
centre of excellence {\em ``Geometric Analysis and Mathematical Physics''} of the 
Academy of Finland and by FP5 EU network contract HPRN-CT-2002-00300. 

\end{document}